\documentclass{jpp}
\usepackage{amsmath, amssymb}
\usepackage{graphicx} %
\usepackage{bm}
\usepackage{tikz, pgfplots}
\usetikzlibrary{decorations.pathreplacing}

\usepgfplotslibrary{colorbrewer}
\usepackage{algorithm,algpseudocode}
\usepackage{siunitx}
\usepackage{bbm}
\usepackage{url}

\usetikzlibrary{decorations,calligraphy}
\DeclareMathSymbol{\sm}{\mathbin}{AMSa}{"39}

\usetikzlibrary{shapes.misc, positioning}

\DeclareMathOperator*{\argmin}{argmin}

\shorttitle{Direct stellarator coil design with global optimization}
\shortauthor{A. Giuliani}

\title{Direct stellarator coil design using global optimization: application to a comprehensive exploration of quasi-axisymmetric devices}

\author{Andrew Giuliani\aff{1}
  \corresp{\email{agiuliani@flatironinstitute.org}},
 }

\affiliation{\aff{1}Flatiron Institute,
162 Fifth Avenue,
New York, NY 10010, USA
}

\begin{document}

\maketitle

\begin{abstract}
Many stellarator coil design problems are plagued by multiple minima, where the locally optimal coil sets can sometimes vary substantially in performance.  As a result, solving a coil design problem a single time with a local optimization algorithm is usually insufficient and better optima likely do exist.  To address this problem, we propose a global optimization algorithm for the design of stellarator coils and outline how to apply box constraints to the physical positions of the coils.  The algorithm has a global exploration phase that searches for interesting regions of design space and is followed by three local optimization algorithms that search in these interesting regions (a ``global-to-local" approach).  The first local algorithm (phase I), following the globalization phase, is based on near-axis expansions and finds stellarator coils that optimize for quasisymmetry in the neighborhood of a magnetic axis.  The second local algorithm (phase II) takes these coil sets and optimizes them for nested flux surfaces and quasisymmetry on a toroidal volume.  The final local algorithm (phase III) polishes these configurations for an accurate approximation of quasisymmetry.  Using our global algorithm, we study the trade-off between coil length, aspect ratio, rotational transform, and quality of quasi-axisymmetry.  The database of stellarators, which comprises approximately 200,000 coil sets, is available online and is called QUASR, for `QUAsi-symmetric Stellarator Repository'.
\end{abstract}

\section{Introduction}
Stellarator coil design is typically framed as an optimization problem, where the objective function targets charged particle confinement, other physics properties, and engineering requirements on the electromagnetic coils. 
Coil design problems can admit multiple local minima \citep{Zhu_2018,genetic} with large variability in performance of the discovered designs \citep{wechsung2022single}.
One remedy of this problem is to use stochastic optimization \citep{silke,wechsung2022single}, which appears to reduce the variability of the discovered minima.
Stochastic optimization of coils has also been explored in \citep{jf1,jf2}.
Another approach is to use global optimization algorithms that attempt to fully explore the range of allowable stellarator coil designs. \citep{silke} combines both remedies by searching for the global optimum of a stochastic objective function.
Genetic algorithms have also been explored in \citep{genetic}.

\begin{figure}
\includegraphics[width=\textwidth]{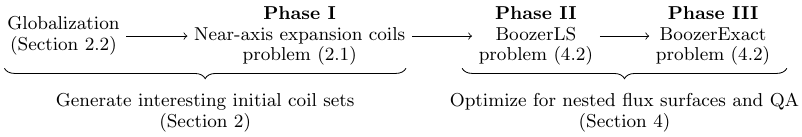}
    \caption{Current workflow wraps phase I (the near-axis expansion algorithm, \citep{nae}) in globalization subroutines \citep{turbo} to generate interesting initial guesses for the subsequent volume QA optimizations in phase II \citep{surfaceopt1} and III \citep{surfaceopt2}.}
    \label{fig:workflow}
\end{figure}

In this work, we propose a novel technique for applying global optimization algorithms to deterministic coil design problems plagued by many local minima.
Our coil design workflow is decomposed into three phases, wrapped in a globalization procedure (Figure \ref{fig:workflow}) to \textit{automatically} design a large database of vacuum field stellarators for various design targets.  
During phase I, the first algorithm finds initial coil sets with nested magnetic surfaces in the neighborhood of a magnetic axis by using the near-axis expansion formalism \citep{nae}.
During phase II, the second algorithm takes these coil sets and expands the region of nested flux surfaces with a good approximation of quasi-axisymmetry (QA) \citep{surfaceopt2}.
Finally, during phase III, the third algorithm polishes these coil sets for precise QA \citep{surfaceopt1}. 

We compare two approaches to globalization.  The first, somewhat naive approach attempts to find a global minimum of the objective by perturbing initial guesses provided to a local optimization algorithm.  The disadvantage is that if the perturbation is too large, then the optimizer might be sent to an uninteresting region of parameter space.  Conversely, if the perturbation is too small, the optimizer might not sufficiently explore the design space.
Despite this downside, it can yield good results with some tuning \citep{wechsung2022single}.
A second, much less ad-hoc approach, relies on the global optimization algorithm called TuRBO \citep{turbo}.
For this algorithm, the user must provide lower and upper bounds on the design variables, known as box constraints.
This algorithm has been used before in coil design problems \citep{silke}, but our approach is notably different as we reformulate our optimization problem to accept constraints on the geometry of the coils in physical, rather than Fourier space.

Since we are generating a large data set of stellarators, we also study the trade-offs between multiple competing design targets, which is the aim of multi-objective optimization.  Recently, multi-objective optimization was applied to stellarator coil design \citep{bindel2023understanding}, where a first order continuation algorithm to construct a local Pareto front was developed.
The algorithm was based on a Taylor expansion of the optimality condition that points on the Pareto front satisfy.
In this work, we do not use this continuation approach but nevertheless study the trade-offs in stellarator coil design using our database.

To summarize our contributions, we propose a simple way to incorporate box constraints on the physical positions of stellarator coils so that global optimization algorithms can be applied to the first phase of the coil design procedure.
Following this first phase are two volume QA algorithms which target nested flux surfaces and precise QA.  Scanning over various physics design targets, we have compiled a comprehensive database of approximately 200,000 stellarator devices, and examined the trade-off between competing design objectives (quality of quasisymmetry, total coil length, rotational transform, and device aspect ratio).
A subset of the devices have comparable quasisymmetry to highly optimized configurations in \citep{surfaceopt2}.

We would also like to point out some limitations of this work.  
First, the globalization is \textit{only} applied to the near-axis coil design algorithm, and not to the volume QA phases of coil design.  Because of this, we may miss out on performant stellarators should precise first-order near-axis QA in phase I not correlate with precise volume QA in phases II and III.
Second, this database only contains curl-free stellarators with optimized quasi-axisymmetry, though our algorithms are generic and apply to other flavors of quasisymmetry.  These vacuum-field devices are can also useful as they can be used as initial guesses during an optimization as plasma pressure is progressively increased \citep{curlfree}. 
Third, there may be duplicate devices in the database and possible mechanisms for this are outlined in Section \ref{sec:volume_database}.
Finally, there are specialized algorithms for visualizing the Pareto front, which we do not use here.

\section{Phase I: stellarators optimized for quasisymmetry on the magnetic axis} \label{sec:opt1}
The goal of the first optimization problem in phase I is to find an initial set of coils that has nested flux surfaces and produces a good approximation of quasisymmetry, at least locally to the magnetic axis.
To this end, we recall here the optimization problem in \citep{nae} that computes coils based on near-axis expansions:
\begin{subequations}\label{eq:optprob1}
\begin{align}
    \min_{\mathbf c, \mathbf I, \mathbf a , \bm \sigma , \iota_a, \overline{\eta}} & ~ \hat f_{\text{axis}}(\mathbf c, \mathbf I, \mathbf a, \bm \sigma, \overline{\eta}, \iota_a) \label{eq:optprob1:J}\\
    \text{subject to } \mathbf{g}(\mathbf a, \bm \sigma, \overline{\eta}, \iota_a) &= 0   \text{ on axis,}  \label{eq:optprob1:2}\\
    \iota_a &= \iota_{\text{target}}, \label{eq:optprob1:3}\\
    L_i &= L_{\text{target}}, \label{eq:optprob1:4}\\
    \kappa_i &\leq \kappa_{\max}, ~ i = 1, \ldots, N_c, \label{eq:optprob1:5}\\
    \frac{1}{L_i}\int_{\bm\Gamma^{(i)}} \kappa_i^2 ~dl &\leq \kappa_{\mathrm{msc}}, ~ i = 1, \ldots, N_c, \label{eq:optprob1:6}\\
    \| \bm\Gamma^{(i)}- \bm\Gamma^{(\text{axis})} \| &\geq d_{\min} ,\label{eq:optprob1:7}  \\
    \| \bm\Gamma^{(i)}- \bm\Gamma^{(j)} \| &\geq d_{\min} ~ \text{ for } i \neq j,\label{eq:optprob1:8}  \\
    \|\bm \Gamma'^{(i)}\| - L_i &= 0 ~\text{ for } i = 1, \ldots, N_c, \label{eq:optprob1:9} 
\end{align}
\end{subequations}
where 
\begin{equation}
\begin{aligned}
\label{eq:objBFGS}
\hat f_{\text{axis}}(\mathbf c, \mathbf I, \mathbf a, \overline{\eta} , \bm \sigma , \iota_a) &= \int_{\text{axis}}(\| \mathbf B(\mathbf c, \mathbf I,\mathbf a)\| - B_{\text{QS}}( \mathbf a) )^2 dl  \\
&+  \int_{\text{axis}} \|\nabla \mathbf B(\mathbf c, \mathbf I,\mathbf a) - \nabla \mathbf B_{\text{QS}}(\mathbf a, \overline{\eta},  \bm \sigma, \iota_a)\|^2 dl,
\end{aligned}
\end{equation}
and $\mathbf{B}, \nabla\mathbf{B}$ are the magnetic field and its gradient generated by the coils,
$\mathbf B_{\text{QS}}, \nabla \mathbf{B}_{\text{QS}}$ are the magnetic field and its gradient, which is quasisymmetric to first order for the given magnetic axis shape. 
$\bm \Gamma^{(\text{axis})}, \bm \Gamma^{(i)} \in \mathbb R^3$ are respectively the vector of the magnetic axis and $i$th coil's Cartesian coordinates.
 The vector of coil degrees of freedom are stored in $\mathbf{c} \in \mathbb{R}^{3N_c(2N_{f,c}+1)}$. 
 The $x$, $y$, $z$ positions of the $N_c$ base modular coils are each represented using a Fourier series with $N_{f,c}$ modes \citep{Zhu_2018}.
 The electromagnetic coils also have an associated current stored in the vector $\mathbf I \in \mathbb R^{N_c}$.  The full device is obtained by applying discrete rotational symmetry with $n_{\text{fp}}$ field periods and stellarator symmetry to $N_c$ base modular coils. 
 The length of the $i$th coil is $L_i = \int_0^1 \|\bm \Gamma'^{(i)}(\theta)\|~d\theta$.
The stellarator symmetric magnetic axis is represented using a Fourier series of the $R$ and $Z$ positions of the magnetic axis, represented in cylindrical coordinates.
In total, this results in $2N_{f,a}+1$ degrees of freedom associated to magnetic axis, stored in $\mathbf{a}\in \mathbb{R}^{2N_{f,a}}$, where the dimensionality is one less than the number of Fourier coefficients of the axis since we require that the mean radius of the magnetic axis be $\SI{1}{\meter}$.
Finally, $\iota_a \in \mathbb R$ is the on-axis rotational transform, and $\overline{\eta} \in \mathbb R \backslash \{0\}$ is a scalar that describes how much that magnetic field strength varies on surfaces in the neighborhood of the axis.
 Given $\mathbf{a}$, and $\overline{\eta}$, the vector $\bm \sigma \in \mathbb{R}^{2N_{f,a}}$ and on-axis rotational transform $\iota_a$ are fully determined via the constraint $\mathbf g$ in \eqref{eq:optprob1:2}, which is the periodic Ricatti equation
 \begin{equation}\label{eq:ricatti}
\frac{d\sigma}{d\varphi} + \iota\left[ \frac{\overline{\eta}}{\kappa^4} + 1 + \sigma^2 \right] + 2\tau \frac{G_0}{B_0} \frac{\overline \eta^2}{\kappa^2} =0,
 \end{equation}
discretized using a Fourier collocation method \citep{nae}, where $\kappa, \tau$ are respectively the curvature and torsion on the magnetic axis and $B_0$ is the field strength on axis.  Other variables in \eqref{eq:ricatti} like $\varphi, \sigma, G_0$ are defined in \citep{landreman2018direct}.
 The data $\iota_a$, $\overline{\eta}$, $\bm \sigma$ fully define the gradient of the target quasi-symmetric magnetic field on axis.
After solving \eqref{eq:ricatti}, then we can evaluate the near-axis gradient $\nabla \mathbf B_{\text{QS}}$.

Using the implicit function theorem, we minimize the reduced objective 
\begin{equation}\label{eq:nae_obj}
f_{\text{axis}}(\mathbf c, \mathbf I, \mathbf a, \overline{\eta}) = \hat f_{\text{axis}}(\mathbf c, \mathbf I, \mathbf a, \overline{\eta}, \bm \sigma(\mathbf a, \overline{\eta}), \iota_a(\mathbf a, \overline{\eta}))
\end{equation}
by eliminating $ \bm \sigma , \iota_a$ via the constraint \eqref{eq:optprob1:2}. Thus, this constraint is satisfied exactly.
 The constraint in \eqref{eq:optprob1:3} ensures that a stellarator is found with the target on-axis rotational transform $\iota_{\text{target}}$.
 The constraint in \eqref{eq:optprob1:4} ensures that each electromagnetic coil has the same length $L_{\text{target}}$.  
 The constraints in \eqref{eq:optprob1:5} and \eqref{eq:optprob1:6} ensure that the maximum curvature and mean squared curvature do not exceed $\kappa_{\max}$, $\kappa_{\text{msc}}$, respectively.
The constraints in \eqref{eq:optprob1:7}, \eqref{eq:optprob1:8} prevent the inter-coil and coil-to-axis distances from decreasing below $d_{\min}$.
Finally, the constraint in \eqref{eq:optprob1:9} ensures that the coil incremental arclength stays uniform.
Constraints \eqref{eq:optprob1:3}-\eqref{eq:optprob1:9} are enforced using a penalty method, and are satisfied to $0.1\%$ precision.
For each of these constraints, a quadratic penalty term appears in the objective function.  For example, the quadratic penalty associated to the equality constraint \eqref{eq:optprob1:3} is $\frac{1}{2}(\iota_a - \iota_{\text{target}})^2$.
To ensure that the constraints are satisfied accurately enough, BFGS is run multiple times with a fixed computational budget.  At the end of BFGS each run, the weight in front of the quadratic penalty was increased by a factor of 10 if the constraint is violated by more than 0.1\%.

A local minimum of \eqref{eq:optprob1} is a set of electromagnetic coils that produce a magnetic field that is close to the quasisymmetric magnetic field on a magnetic axis.
This optimization problem was solved using local quasi-Newton optimization methods in \citep{nae, wechsung2022single}, where the analytical gradient of the reduced objective $f_{\text{axis}}$ was obtained using both forward sensitivities and adjoint approaches.  
We always use the Broyden-Fletcher-Goldfarb-Shannon (BFGS) quasi-Newton method in phase I of the workflow.

In the next two sections, we both illustrate the need for globalization when solving \eqref{eq:optprob1} and outline two different techniques to accomplish this.
\subsection{Naive globalization of phase I}\label{sec:naive}
A first naive approach to globalization is to perturb the geometry and currents of flat electromagnetic coils, as well as the parameters of the magnetic axis to generate a set of initial starting points.
These starting points are used to initialize Phase I, where the local gradient-based optimization algorithm BFGS is used to find a local minimum of the objective in the neighborhood of the perturbed initial guess.
The physics and engineering quantities that we target are $\iota_{\text{target}}=0.9$ using $n_{\text{fp}}, n_{\text{coils per hp}}=2$, $d_{\min}=\SI{0.1}{\meter}$, $\kappa_{\max}=\SI{5}{\meter^{-1}}$ $\kappa_{\text{msc}}=\SI{5}{{\meter}^{-2}}$, the target coil length $L_{\text{target}}$ is varied between 40 and 80 meters. 
For each value of coil length, we solve \eqref{eq:optprob1} with distinct initial guesses 16 times.
The initial guesses are obtained by perturbing the Fourier coefficients of initially flat coils with normally distributed noise with zero mean and standard deviation $\epsilon$.
One has substantial freedom to decide many Fourier coefficients to perturb, as well as the standard deviation of the perturbation and how it decays.
If $\epsilon$ is too large then the local optimizer might be sent into an uninteresting region of coil parameter space.  If $\epsilon$ is too small, and it might not fully explore the set of feasible coil designs.
For the experiment here, we perturb the currents, $\overline{\eta}$,  and only the first two Fourier harmonics of the coils, and magnetic axis.   Extensive (but tedious and computationally intensive) tuning revealed that the standard deviation of the noise that discovered devices with lowest quasi-symmetry error was $\epsilon = \SI{0.01}{\meter}$.  We find that the quality of the minima found depends strongly on the choice of $\epsilon$, the number of Fourier modes that are perturbed, and that there is no way to know a priori that this value works well for other stellarator designs.  
\begin{figure}
    \centering
    \includegraphics[width=\textwidth]{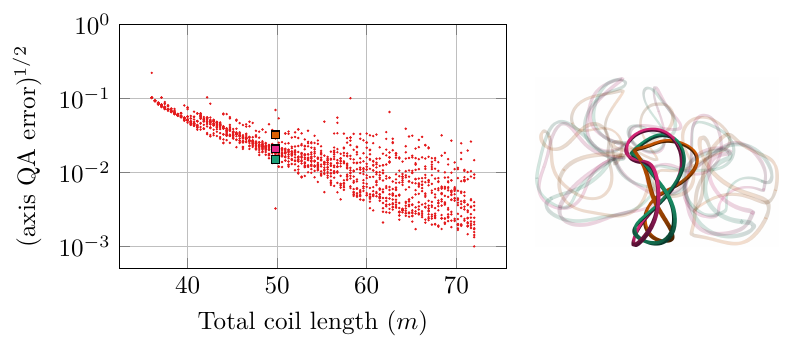}
    \caption{On the left, we plot the on-axis quasisymmetry error with respect to the total coil length used, where we observe that there are multiple minima. On the right, we plot four distinct local minima when the total coil length is $\SI{49.81}{\meter}$, where the colors of the coils correspond to those of the squares on the left.  The coil sets' associated magnetic axes are not plotted to avoid cluttering in the image.}
    \label{fig:motivation}
\end{figure}

In Figure \ref{fig:motivation}, we plot the trade off between the on-axis quasisymmetry error and total coil length.  We find as expected that lower quasisymmetry errors are attainable with longer coils.
It is clear that making no attempt at globalization to solve this optimization problem might result in a sub-optimal coil set.
The problem of multiple minima becomes more pronounced for devices with longer coil lengths because there are more, intricate ways for the coils to arrange themselves.
Finally, at around $\SI{50}{\meter}$, there is a device that appears as an outlier with remarkably low near-axis quasisymmetry error.
This hints that the naive approach is missing interesting solution branches.

\subsection{TuRBO globalization of phase I}\label{sec:g2l}
An alternative to the naive approach is to use a global optimization algorithm.  The coil set it finds is then given as the starting point in Phase I to a local gradient-based optimization algorithm BFGS.
This global-to-local approach was explored in \citep{silke}, where stochastic global optimization of coil geometries was done using the DTuRBO algorithm \citep{dturbo}, an extension of the gradient-free global optimization algorithm TuRBO \citep{turbo} that takes into account gradients.   
The TuRBO algorithm solves the minimization problem
\begin{equation}
\begin{aligned}\label{eq:global_problem}
\min_{\mathbf x \in \mathbb R^N} &f(\mathbf x)\\
\ell_i \leq x_i& \leq u_i ~ \text{ for } i = 1 \hdots  N,
\end{aligned}
\end{equation}
where $N$ is the dimension of the optimization problem, $\ell_i, u_i$ are respectively lower and upper bounds on the components of the control variables.
That is, the algorithm attempts to find the global minimum $f(\mathbf x)$ on an $N$-dimensional hyper-rectangle.  The TuRBO algorithm completes a preliminary exploration phase by quasi-uniformly sampling the hyper-rectangle to find interesting areas of parameter space.  Then, it uses trust-region based Bayesian optimization algorithms to further refine the solution.

It is difficult to define sensible lower and upper bounds in \eqref{eq:global_problem} on the unknowns when they are the Fourier coefficients of the coils.  There should be some decay in the size of the boxes with increasing mode number, but it unclear how this decay should be chosen.  
In \citep{silke}, manufacturing errors were projected onto a Fourier basis, from which reasonable bounds on the Fourier coefficients were inferred.
In this work, we propose a different way of determining the box constraints for deterministic coil design problems, though, our technique is generic enough to be used in the stochastic context as well.

\subsubsection*{Bounding boxes}
It is straightforward to constrain the coil currents to $-1 \leq  \mu_0 I_i  \leq 1$,  where $\mu_0 = 4\pi \times 10^{-7}$ is the magnetic constant.  The parameter $\overline \eta$ can also be constrained by $0\leq \overline{\eta} \leq 2$.

\begin{figure}
    \centering
    \includegraphics[width=\textwidth]{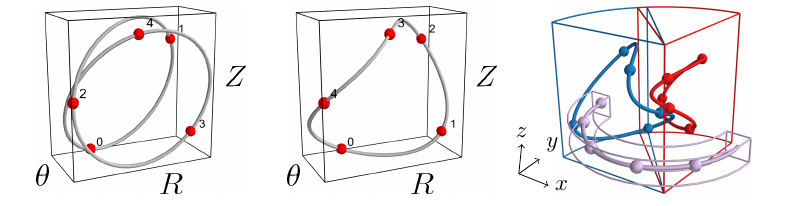}
    
    \caption{The randomly generated coil in cylindrical coordinates on the left is too complex.  Re-indexing the nodes by ordering them about their barycenter in the $RZ$ plane, the coil untangles in the center figure.  On the right, a set of coils, the magnetic axis, and their associated bounded boxes in physical space are illustrated.  The full stellarator is not shown, but can be obtained by applying symmetries.}
    \label{fig:reindexed}
\end{figure}

We still use a Fourier series to represent the coils and axis, but change the degrees of freedom from Fourier coefficients to spatial coordinates, linked to each other by the discrete Fourier transform.
In this way, physically meaningful box constraints can be provided to TuRBO.
These positions, or anchor points, in cylindrical coordinates, $(r_i, \theta_i, z_i)$, are constrained to the box $[0, 1 + R_{\text{minor}}] \times [\theta_i-\Delta \theta/2, \theta_i + \Delta \theta/2] \times [-R_{\text{minor}}, R_{\text{minor}}]$, where $R_{\text{minor}} = L_{\text{target}}/2\pi$, and $\Delta \theta = \rho (2\pi/2n_{\text{fp}} n_{\text{coils per hp} })$, $\theta_i = (2\pi/2n_{\text{fp}}n_{\text{coils per hp} })(i+1/2)$, $R_{\text{minor}}$ is the radius of the perfectly circular coil with length $L_{\text{target}}$, $\Delta \theta$ defines a cylindrical sector that the coil can occupy, and the $\rho$ multiplier ensures that the coil bounding boxes overlap somewhat.
Randomly sampling $2N_{f,c}+1$ points on the box will frequently generate coils with complex geometries (Figure \ref{fig:reindexed}, left).
These geometries are not particularly useful and we would like TuRBO to avoid wasting time in these uninteresting areas of parameter space.
These configurations can be avoided, while still using the same anchor points, by re-indexing them so that the points are ordered counterclockwise about their barycenter in the $RZ$ plane (Figure \ref{fig:reindexed}, center). This unravels the coils. 
Finally, the these coil coordinates are converted to Cartesian coordinates, projected onto a Fourier basis, then used to evaluate the near-axis objective \eqref{eq:nae_obj}.

We also generate $(R_i, \phi_i, Z_i)$ positions, or anchor points, for the magnetic axis, where $(R_i, Z_i)$ are constrained to the box $[1-1/(1+n^2_{\text{fp}}), 1+ 1/(1+n^2_{\text{fp}})] \times [-0.2, 0.2]$, and $\phi_i$ is a uniformly spaced cylindrical angle in $2\pi/n_{\text{fp}}$. 
The bounding boxes for $R_i$ and $Z_i$ are informed by the fact that quasi-axisymmetric magnetic axes have zero helicity. Helicity is an integer associated to the axis geometry that measures how many poloidal transits the axis normal vector makes as the axis traces one toroidal revolution.  
It was shown in \citep{rodriguez2022phases} that magnetic axes of the form
\begin{align*}
    R(\phi) &= 1+a\cos(n_{\text{fp}}\phi), \\
    Z(\phi) &= b\sin(n_{\text{fp}}\phi),
\end{align*}
have zero helicity when $a < 1/(1+n^2_{\text{fp}})$. 
The situation is more delicate if additional Fourier harmonics are used to represent the axis (as we do here), but in the case of a single dominant harmonic, this bound on $a$ is a well-informed estimate.
The box associated to $Z_i$ was chosen to prevent large excursions from the $Z=0$ plane, though the size of this box was somewhat arbitrary and other values might have been used.

There is one final detail that ensures that the resulting axis curve is stellarator symmetric.
After sampling $N_{f,a}+1$ points from the bounding box associated to the radial coordinates $R_i$, they are arranged in an $2N_{f,a}+1$-sized array as follows:
$$
[R_0-s, R_1-s, \hdots, R_{N_{f,a}}-s, R_{N_{f,a}}-s, \hdots, R_1-s],
$$
where $s$ is a shift to ensure the array has mean 1.  After the shift, the radial positions of the magnetic axis may lie slightly outside the original bounding box.
Similarly, we sample $N_{f,a}$ points from the bounding box associated to the vertical coordinates $Z_i$, and arrange them in a $2N_{f,a}+1$-sized array as follows:
$$
[0, Z_1, \hdots, Z_{N_{f,a}}, -Z_{N_{f,a}}, \hdots, -Z_1].
$$
The stellarator symmetric $R$ and $Z$ harmonics of the magnetic axis are obtained by discrete Fourier transform of these arrays and provided to the near-axis objective \eqref{eq:nae_obj}.

The bounding boxes of different coils can overlap each other and all the bounding boxes of the coils necessarily overlap the one of the magnetic axis, since we do not search for windowpane coils.
Thus, it is possible for the generated coils to be linked, or for the coils to not be linked with the magnetic axis.  
In addition, it is still possible that the magnetic axis sampled from the bounding boxes to have nonzero helicity.
It is for this reason that we add penalty terms to the objective $f_{\text{axis}}$ that avoid these undesirable configurations:
\begin{equation}\label{eq:objTuRBO}
f(\mathbf c , \mathbf I, \mathbf a, \overline{\eta}) = f_{\text{axis}}(\mathbf c ,  \mathbf I, \mathbf a , \overline{\eta}) + w\biggl[\sum_{i}\sum_{j>i} l_{i,j}(\mathbf{c}) +  \sum_{i}|l_{i,\text{axis}}(\mathbf{c}, \mathbf a)-1| +  h(\mathbf a)\biggr],
\end{equation}
where we abuse notation here and use $\mathbf c, \mathbf a$ to represent the vector of anchor points on the coils and axis, respectively.   These positions can be converted to the Fourier coefficients by using a discrete Fourier transform.  The value $l_{i,j}$ is the absolute value of the linking number between coils $i$ and $j$, and $l_{i, \text{axis}}$ is the absolute value of the linking number between coil $i$ and the magnetic axis.  
For the linking number calculation, we use the implementation in \citep{bertolazzi2019efficient} since it returns a linking number with certified accuracy by tracking rounding errors in the calculation.
Finally, $h(\mathbf{a})$ is the absolute value of the axis helicity. 
The weight $w=10^7$ is chosen to make the terms that it multiplies greater than $f_{\text{axis}}$.
The penalty terms multiplied by $w$ are zero when the design variables describe a stellarator with unlinked coils that each wrap once poloidally around a magnetic axis with zero helicity.
Since the objective is no longer differentiable due to the linking number and helicity calculation, we use TuRBO rather than DTuRBO.

We also designed stellarators using an axis bounding box independent of $n_{\text{fp}}$, $(R_i, Z_i) \in [0.85, 1.15] \times [-0.15, 0.15]$, and neglecting to include the helicity penalty.  In a postprocessing step, devices that had an axis with nonzero helicity were discarded.  This approach risks discarding devices and wasting computational resources but it is capable of producing comparable results.

\subsection{Illustration of TuRBO globalization followed by phase I}\label{sec:turbo_illustration}
To illustrate the global-to-local procedure,  we examine the TuRBO globalization procedure followed by the phase I optimization for a single stellarator.
We use a batch size of 100 and restrict TuRBO to a budget of 15,000 function evaluations, of which 1,000 are used in the initial exploration phase.   TuRBO's internal Gaussian process model is only trained using the most recent 1,000 function evaluations for speed.
Finally, the underlying Fourier discretization uses $N_{f,c}=N_{f,a}=2$ for the coils and axis during this global exploration phase.
The coil set and magnetic axis from the global search are used as the initial guess to the BFGS algorithm to polish the solution.
During this phase, the number of Fourier modes used is increased to $N_{f,c}=N_{f,a}=6$.
Before BFGS obtains a good enough approximation to the inverse Hessian, especially during the first few iterations, the line search might accept a large enough step such that the coils become interlocked or unlinked with the magnetic axis.  To avoid this, we use a linking number calculation to trigger the line search.

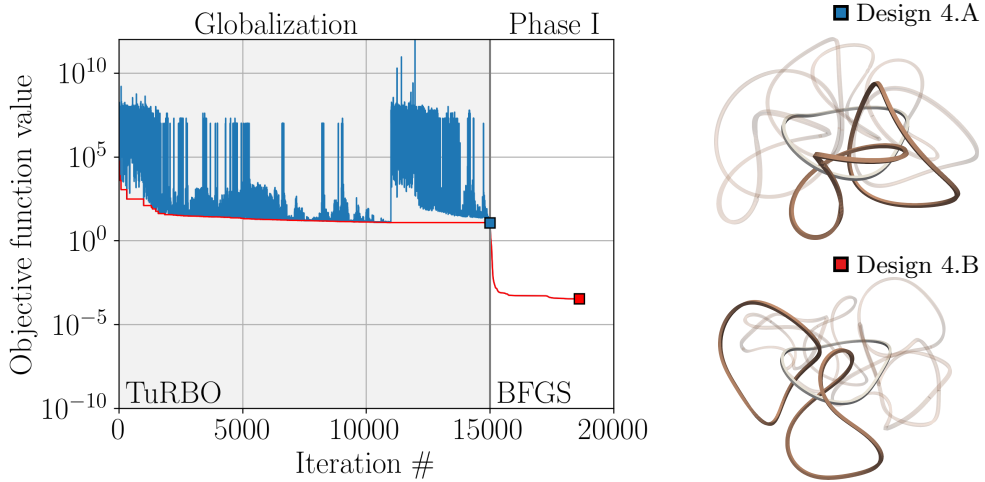
\begin{figure}
    \centering
    \includegraphics[width=\textwidth]{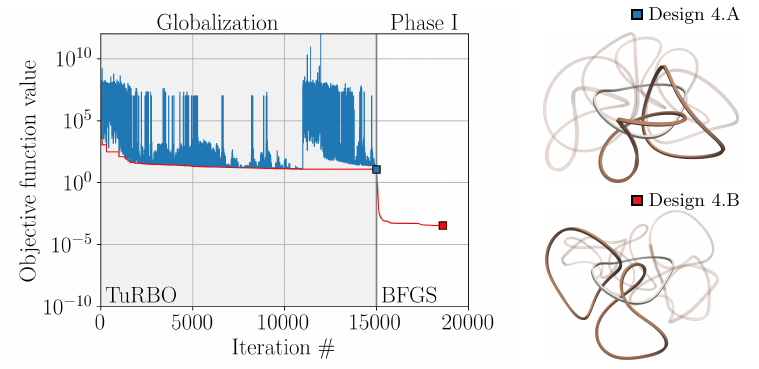}
    \caption[]{The value of the objective function when applying TuRBO then BFGS.  For the TuRBO phase, the objective values are evaluations of \eqref{eq:objTuRBO} , which includes the non-differentiable penalties and the `iteration \#' is actually number of function evaluations. For the BFGS phase, the objective values are evaluations of \eqref{eq:objBFGS}, and `iteration \#' refers to the objective function values after an accepted line search.   The red line is the best function evaluation so far.  
    On the right, we plot the best stellarator found by TuRBO, labeled \begin{tikzpicture}
         \draw[color=black, fill=Paired-B, line width=1pt]  (0,0) rectangle ++ (0.16,  0.16);
    \end{tikzpicture} Design \ref{fig:TuRBO_then_BFGS}.A, provided as the initial guess to BFGS in phase I.  Then, the stellarator found by BFGS after increasing the number Fourier modes is given by \begin{tikzpicture}
         \draw[color=black, fill=Paired-F, line width=1pt]  (0,0) rectangle ++ (0.16,  0.16);
    \end{tikzpicture}  Design \ref{fig:TuRBO_then_BFGS}.B.}
    \label{fig:TuRBO_then_BFGS}
\end{figure}

In Figure \ref{fig:TuRBO_then_BFGS}  we provide the objective function evaluations over the course the TuRBO globalization then phase I of Figure \ref{fig:workflow}. 
The TuRBO phase of the run has large variations in the objective function, varying over several orders of magnitude. 
When the objective is larger than $10^7$, this likely corresponds to a device that has an axis with nonzero helicity, linked coils, or coils unlinked with the axis.
TuRBO quickly finds an interesting region of parameter space and it could have used a smaller budget.  
The notable jump in the objective function value around 11,000 function evaluations occurs when TuRBO no longer can make progress in the trust region, so it is discarded and another one is initialized.
The best configuration found by TuRBO is provided and labeled Design \ref{fig:TuRBO_then_BFGS}.A.  
The coil set at the end of Phase I (Design \ref{fig:TuRBO_then_BFGS}.B) has changed substantially from those obtained after the TuRBO phase.

\subsection{Comparison of devices from TuRBO and naive globalization after phase I}
\begin{figure}
    \centering
    \includegraphics[width=\textwidth]{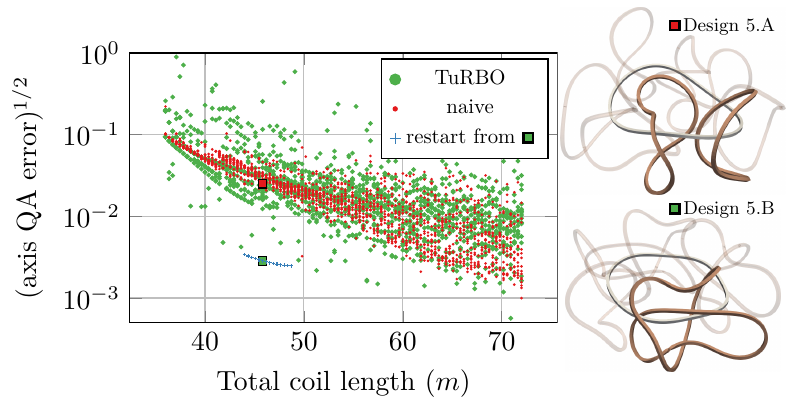}
    \caption[]{On the left, we provide the trade-off between axis quasisymmetry error and total coil length used for local minima found using the naive approach and TuRBO with $\iota=0.9$, and $n_{\text{fp}}, n_{\text{coils per hp}}=2$.  
    We have plotted two coil sets, called Designs \ref{fig:l2g_axis}.A and \ref{fig:l2g_axis}.B, corresponding respectively to the red (\begin{tikzpicture}
        \draw[color=black, fill=Set1-A, line width=1pt]  (6.6,  3.2) rectangle ++ (0.16,  0.16);
    \end{tikzpicture}), and green (\begin{tikzpicture}
        \draw[color=black, fill=Set1-C, line width=1pt]  (6.6,  3.2) rectangle ++ (0.16,  0.16);
    \end{tikzpicture}) square markers.  
    Using Design \ref{fig:l2g_axis}.B as an initial guess, we solve \eqref{eq:optprob1} for different target coil lengths, resulting in the devices corresponding to the blue crosses  (\begin{tikzpicture}
        \draw[Set1-B, line width=1pt]  (0,  0) -- ++ (0.16,  0.);
        \draw[Set1-B, line width=1pt]  (0.08,  0) -- ++ (0.0,  0.08);
        \draw[Set1-B, line width=1pt]  (0.08,  0) -- ++ (0.0, -0.08);
    \end{tikzpicture}). 
    The axis QA error refers to the terms in \eqref{eq:objBFGS}.
    }
    \label{fig:l2g_axis}
\end{figure}

In this section, we compare the performance of TuRBO, with the more naive approach by solving \eqref{eq:optprob1} for many target coil length values when $\iota=0.9$, $n_{\text{fp}}, n_{\text{coils per hp}}=2$.  To this end, we execute the TuRBO globalization followed by phase I, as illustrated in Section \ref{sec:turbo_illustration} and Figure \ref{fig:TuRBO_then_BFGS}.  In both naive and TuRBO approaches, the workflow is executed 16 times per target coil length.  
The quasisymmetry error of the resulting configurations are provided in Figure \ref{fig:l2g_axis}.
TuRBO finds devices that outperform those found by the naive approach, and we plot two devices from each algorithm, called Designs \ref{fig:l2g_axis}.A and \ref{fig:l2g_axis}.B.  We find that Design \ref{fig:l2g_axis}.B is from a genuinely different solution branch.  To drive this home, we use it as an initial guess, we solve \eqref{eq:optprob1} for multiple different target coil lengths.   This solution branch persists on both sides, and was evidently even missed by TuRBO.
This highlights the difficulty of finding global minima.

With very modest input from the practitioner, TuRBO does a good job of localizing performant minima.
We find that the choice of the overlap factor $\rho$ only modestly affected the quality of the minimizers found.  
In contrast, one has to decide on how many Fourier modes to perturb and the size of each perturbation to use with the naive approach.  This can have a large effect on the quality of the minimizers found.

\section{The near-axis coil sets obtained after phase I}
We solve \eqref{eq:optprob1} multiple times with the combinations of following design targets
\begin{equation}
\begin{aligned}
    \iota_{\text{target}} &= 0.1, 0.2, 0.3, \hdots, 0.9,\\
    L_{\text{target}} &= 4.5, 4.75, 5, \hdots, 8.5, 8.75, 9,\\
    n_{\text{coils per hp}} &= 1, 2, 3, 4, \hdots, 13,\\
    n_{\text{fp}} &= 1, 2, 3, 4, 5,
\end{aligned}
\end{equation}
using both naive and TuRBO globalization approaches, then combining the data sets obtained.  
To avoid unrealistic configurations, we only use combinations of the above design targets such that the total coil length used $2n_{\text{fp}}n_{\text{coils per hp}}L_{\text{target}} \leq \SI{120}{\meter}$.
For all devices and physics targets, we use $d_{\min}=\SI{0.1}{\meter}$, $\kappa_{\max}=\SI{5}{\meter^{-1}}$ $\kappa_{\text{msc}}=\SI{5}{{\meter}^{-2}}$.
\begin{figure}
    \centering
    \includegraphics[width=\textwidth]{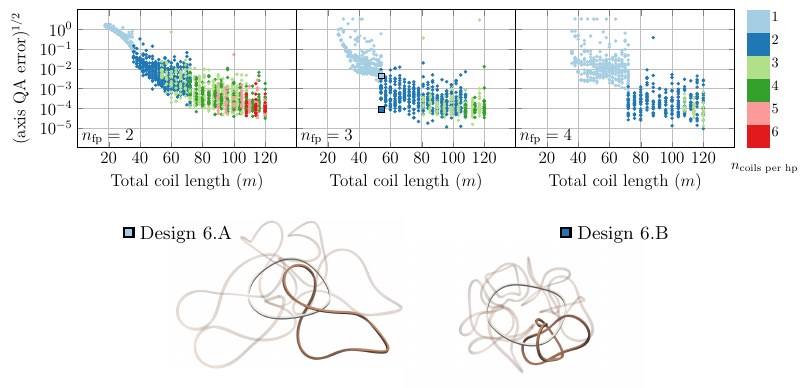}
    \caption{The near axis QA error with respect to total coil length used when $\iota=0.5$, each point on the scatter plot corresponds to a different device and each panel corresponds to a different number of field periods, and the color refers to the number of coils per half period used.  
    Below, we plot two configurations (Designs \ref{fig:full_axis_database}.A and \ref{fig:full_axis_database}.B) that use the same total coil length but different number of coils, 1 and 2 coils per half period, respectively.    
    The axis QA error refers to the terms in \eqref{eq:objBFGS}.
    }
    \label{fig:full_axis_database}
\end{figure}
In Figure \ref{fig:full_axis_database}, we provide the on-axis quasi-symmetry error for a subset of the stellarators, where we observe the expected trade-off between axis quasisymmetry and total coil length: as the total coil length gets longer, the minimum attainable quasisymmetry error decreases and levels off.
If one fixes the total coil length used, but increases the number of coils per half-period, this can drastically improve the on-axis quasisymmetry error.  This is illustrated in Designs A and B plotted on the bottom row of Figure \ref{fig:full_axis_database}.  Both stellarators use the same length of coil ($\SI{54}{\meter}$), but have 1 and 2 coils per half-period respectively, where shorter, more numerous coils are preferred.
The coils in Design B also interleave with one another, in a similar fashion to the coils in Figure \ref{fig:l2g_axis}.
It appears that we can find configurations with a good approximation of on-axis quasi-symmetry for all values of $n_{\text{fp}}$.  Moreover, for a fixed coil length increasing the number of field periods seems to improve the on-axis quasisymmetry error, though the coil-to-axis distance decreases.
As we will observe in the next section, this observation does not bear out when optimizing for quasisymmetry on a volume.  

The generated coil sets can produce nested flux surfaces, but sometimes only over a small volume in the neighborhood of the magnetic axis, making it a good set of initial conditions for the next phase of coil optimization.

\section{Phase II and III: stellarators optimized for quasisymmetry on a volume}\label{sec:opt2}
Taking coil sets optimized for near-axis quasisymmetry obtained during phase I, we now optimize for quasisymmetry on a volume with varying aspect ratios during phases II and III.    
The objective that we minimize is the sum of the average (normalized) quasi-axisymmetry error and the Boozer residuals on $N_s$ surfaces parametrized in Boozer coordinates $\varphi, \theta$:
\begin{scriptsize}
\begin{equation}
\begin{aligned}
\hat{f}_{\text{surface}}(\mathbf{c}, \mathbf I,\mathbf{s}_1, \iota_1, G_1, \hdots, \mathbf s_{N_s}, \iota_{N_s}, G_{N_s}) &:= \frac{1}{N_s}\sum^{N_s}_{k=1} \biggl\{\frac{ \int_0^1 \int_0^{1/n_{\text{fp}}} B_{\text{non-QA}}(\mathbf{c}, \mathbf I, \mathbf{s}_k, \varphi, \theta)^2 ~\|\mathbf n(\mathbf s_k, \varphi, \theta)\|~d\varphi ~d\theta }{  \int_0^1 \int_0^{1/n_{\text{fp}}} B_{\text{QA}}(\mathbf{c}, \mathbf I, \mathbf{s}_k, \theta)^2 ~\|\mathbf n(\mathbf s_k, \varphi, \theta)\|~d\varphi ~d\theta} \\
&+ \frac 12 w_r\int_{0}^1 \int_0^{1/n_{\text{fp}}} \|\mathbf{r}(\mathbf{s}_k, \iota_k, G_k, \mathbf c, \mathbf I, \varphi, \theta)\|^2~d\varphi ~d\theta \biggr\},
\end{aligned}\label{eq:objf4}
\end{equation}
\end{scriptsize}
where Cartesian coordinates of points on the $k$th surface $\bm \Sigma(\mathbf s_k, \varphi, \theta) \in \mathbb R^3$ are expressed using the Fourier representation in \citep{surfaceopt1}, and has $n_{s}$ degrees of freedom associated to it, stored in $\mathbf s_k \in \mathbb R^{n_{s}}$.  In the objective function above, the surface normal is $\mathbf n(\mathbf s_k, \varphi, \theta) = 
\frac{\partial \bm\Sigma}{\partial \varphi}(\mathbf s_k,\varphi,\theta) \times \frac{\partial \bm\Sigma}{\partial \theta}(\mathbf s_k,\varphi,\theta)$,  and $w_r>0$ is a weighting parameter in front of a PDE residual, described below.  
The electromagnetic coils also have an associated current stored in the vector $\mathbf I \in \mathbb R^{N_c-1}$, where the dimensionality is one less than the number of coils because the first coil's current is fixed, preventing the currents from approaching zero. 
Analogously to the previous phase, the vector of coil degrees of freedom are stored in $\mathbf{c} \in \mathbb{R}^{3N_c(2N_{f,a}+1)}$. 
The $x$, $y$, $z$ positions of the $N_c$ base modular coils are each represented using a Fourier series with $N_{f,c}$ modes.
The Boozer residual $\mathbf r$, and scalars $G_k$ and $\iota_k$ are described in more detail below.

The terms in the quasi-axisymmetry error ratio of \eqref{eq:objf4}, introduced in \citep{surfaceopt1}, are given by
$$
B_{\text{QA}}(\mathbf{c}, \mathbf I, \mathbf{s}_k, \theta) = \frac{\int^{1/n_{\text{fp}}}_{0} B(\bm \Sigma(\mathbf s_k,\varphi,\theta), \mathbf c, \mathbf I)~\| \mathbf n(\mathbf s_k, \varphi, \theta)\|~d\varphi}{\int^{1/n_{\text{fp}}}_{0} ~ \|\mathbf n(\mathbf s_k, \varphi, \theta) \|~d\varphi},
$$
with
$$
B_{\text{non-QA}}(\mathbf{c}, \mathbf I, \mathbf{s}_k, \varphi, \theta)  = B_{\text{QA}}(\mathbf{c}, \mathbf I, \mathbf{s}_k, \theta) -B(\mathbf{c}, \mathbf I, \mathbf{s}_k, \varphi, \theta) ,
$$
where $B(\bm \Sigma(\mathbf s_k,\varphi,\theta), \mathbf c, \mathbf I)$ is the field strength evaluated on the $k$th magnetic surface.

The goal is to design a set of coils that solves the following optimization problem:
\begin{subequations}\label{eq:optprob2}
\begin{align}
    \min_{\mathbf c, \mathbf I , \mathbf{s}_1, \iota_1, G_1,\hdots, \mathbf s_{N_s}, \iota_{N_s}, G_{N_s}} & ~ \hat f_{\text{surface}} (\mathbf c, \mathbf I, \: \mathbf{s}_1, \iota_1, G_1, \hdots, \mathbf s_{N_s}, \iota_{N_s}, G_{N_s}) \label{eq:optprob2:J}\\
        \text{subject to }  \mathbf{g}_k(\mathbf{s}_k, \iota_k, G_k, \mathbf c, \mathbf I)=0 &\text{ for $k=1, \ldots, N_s$,}  \label{eq:optprob2:2}\\
    \iota_{\text{target}} - \frac{1}{N_s}\sum_{k=1}^{N_s} \iota_k &= 0, \label{eq:optprob2:3}\\
    R_{\text{major}}(\mathbf s_{N_s}) &= 1, \label{eq:optprob2:4}\\
    \sum_{i = 1}^{N_c} L_i &\leq L_{\max}, \label{eq:optprob2:5}\\
    \kappa_i &\leq \kappa_{\max}, ~ i = 1, \ldots, N_c, \label{eq:optprob2:6}\\
    \frac{1}{L^{(i)}_{c}}\int_{\bm\Gamma^{(i)}} \kappa_i^2 ~dl &\leq \kappa_{\mathrm{msc}}, ~ i = 1, \ldots, N_c, \label{eq:optprob2:7}\\
    \| \bm\Gamma^{(i)}- \bm\Gamma^{(j)} \| &\geq d_{\min} ~ \text{ for } i \neq j,\label{eq:optprob2:8} \\
    \|\bm \Gamma'^{(i)}\| - L^{(i)} &= 0 ~\text{ for } i = 1, \ldots, N_c.
    \label{eq:optprob2:9}
\end{align}
\end{subequations}
The constraint $\mathbf g_k$ is a system of equations that relates the coil and surface degrees of freedom.  This constraint is a spatial discretization that approximates solutions to the system of partial differential equations 
\begin{equation}\label{eq:pde}
\begin{aligned}
    \mathbf r(\mathbf s_k, G_k, \iota_k, \mathbf c, \mathbf I, \varphi, \theta)&=0,\\
    V(\mathbf s_k) - V_{\text{target}, k} &= 0,
\end{aligned}
\end{equation}
where the residual $\mathbf r(\mathbf s_k, G_k, \iota_k, \mathbf c, \mathbf I, \varphi, \theta) \in \mathbb R^3$ is
\small
\begin{equation} \label{eq:residual}
\begin{aligned}
\mathbf r(\mathbf s_k, G_k, \iota_k, &\mathbf c, \mathbf I, \varphi, \theta) := G_k \frac{\mathbf B(\bm \Sigma(\mathbf s_k, \varphi, \theta), \mathbf c, \mathbf I)}{\| \mathbf B (\bm \Sigma(\mathbf s_k,\varphi, \theta), \mathbf c, \mathbf I)\|} \\
&- \|\mathbf B(\bm \Sigma(\mathbf s_k,\varphi, \theta), \mathbf c, \mathbf I)\| \left( \frac{\partial \bm \Sigma(\mathbf s_k,\varphi, \theta)}{\partial \varphi} + \iota_k \frac{\partial \bm \Sigma(\mathbf s_k,\varphi, \theta)}{\partial \theta} \right),
\end{aligned}
\end{equation}
\normalsize
and $V(\mathbf s_k)$ is the volume enclosed by the surface $\bm \Sigma(\mathbf s_k, \varphi, \theta)$ and $V_{\text{target},k}$ is the user-specified target volume of the surface.
The inputs to the PDE are the surface label $V_{\text{target},k}$, and the electromagnetic coils $\mathbf c, \mathbf I$.  The outputs are the surface degrees of freedom $\mathbf s_k$, $G_k$, and rotational transform $\iota_k$.
Note that for a given surface solve, the geometry of the coils $\mathbf c$ and their currents $\mathbf I$ are fixed.

We have devised two approaches to solve \eqref{eq:pde} using a collocation method. 
In both approaches, constraint \eqref{eq:optprob2:2} allows us to compute the surface degrees of freedom $\mathbf s_k, \iota_k, G_k$ in terms of the electromagnetic coils $\mathbf c, \mathbf I$.

The approach used during phase II, called BoozerLS surfaces, is useful for regimes where nested flux surfaces do not exist, and solutions to \eqref{eq:pde} do not exist.  In this case, we search for solutions that satisfy the PDE in a least squares sense.  This results in a bilevel optimization problem, comprising an outer optimization problem over the coil geometries and currents to optimize the device for quasisymmetry.  The inner optimization problem is over the surface degrees of freedom for fixed coil geometry $\mathbf c$ and currents $\mathbf I$, and computes surfaces that solve \eqref{eq:pde} in a least squares sense:
\begin{equation}\label{eq:inner_problem}
\begin{aligned}
\mathbf s_k(\mathbf c, \mathbf I), G_k(\mathbf c, \mathbf I), \iota_k(\mathbf c, \mathbf I) = \argmin_{\widetilde{\mathbf s}_k, \widetilde{G}_k, \widetilde{\iota}_k} &\int_{0}^1\int_0^{1/n_{\text{fp}}} \| \mathbf r(\widetilde{\mathbf s_k}, \widetilde{G}_k, \widetilde{\iota}_k, \varphi, \theta, \mathbf c, \mathbf I) \|^2~d\varphi~d\theta \\
&+ \frac{1}{2}w_{v,k}(V(\widetilde{\mathbf{s}}_k) - V_{\text{target},k})^2,
\end{aligned}
\end{equation}
where the first term in the objective above measures how accurately the PDE is solved and the second term premultiplied by the weight $w_{v,k}$ ensures that we solve for a magnetic surface with target volume $V_{\text{target},k}$.
In this case, \eqref{eq:optprob2:2} is the optimality condition of the inner least squares optimization problem \eqref{eq:inner_problem} solved by BFGS.
 The approach makes the surface solve more robust because we do not require the residual be zero at a set of collocation points and we can use robust line search based algorithms to minimize the PDE residual.  
 That being said, solving this inner optimization problem to sufficient accuracy can be expensive, which is the price paid for robustness.
 
The approach used during phase III, called BoozerExact surfaces, is useful for regimes when nested flux surfaces do exist and we would like to polish the quality of QA.
In this case, \eqref{eq:optprob2:2} is the vector of residual  \eqref{eq:residual} evaluations on a tensor product grid of collocation points in $(\varphi, \theta)$, which we require to be zero.  
This is a stronger requirement, and makes the numerical method more brittle: when nested magnetic surfaces do not exist, Newton's method may fail.  At the expense of robustness, solving the constraint here is much less computationally expensive.

Using the implicit function theorem, we minimize the reduced objective 
$$
f_{\text{surface}}(\mathbf c, \mathbf I) = \hat f_{\text{surface}}(\mathbf c, \mathbf I, \mathbf s_1(\mathbf c, \mathbf I),\iota_1(\mathbf c, \mathbf I), G_1(\mathbf c, \mathbf I), \hdots, \mathbf s_{N_s}(\mathbf c, \mathbf I), \iota_{N_s}(\mathbf c, \mathbf I), G_{N_s}(\mathbf c, \mathbf I))
$$ 
by eliminating $ \mathbf s_k$, $ \iota_k$, $ G_k$ via the constraint \eqref{eq:optprob2:2}.  
Constraint \eqref{eq:optprob2:3} ensures that the mean rotational transform on the volume is the target value $\iota_{\text{target}}$.
Constraint \eqref{eq:optprob2:4} ensures that the sum of coil lengths per half-period is less than $L_{\max}=n_{\text{coils per hp}}L_{\text{target}}$, where $L_{\text{target}}$ is the coil length used in the near-axis optimization problem \eqref{eq:optprob1}.
This is notably different to previously, where each coil was constrained to have the same length $L_{\text{target}}$.
Constraint \eqref{eq:optprob2:5} ensures that the major radius of the outermost surface is 1.
Constraints \eqref{eq:optprob2:6}, and \eqref{eq:optprob2:7} ensure that the maximum coil curvature, and mean squared curvature do not respectively exceed $\kappa_{\max}$ and $\kappa_{\text{msc}}$. %
Constraint \eqref{eq:optprob2:7} ensures that the minimum intercoil distance does not decrease below $d_{\min}$.
Finally constraint \eqref{eq:optprob2:9} ensures that the coils have uniform incremental arclength.
Constraints \eqref{eq:optprob2:3}-\eqref{eq:optprob2:9} are enforced using a penalty method, and are satisfied to $0.1\%$ precision.
We always use the BFGS quasi-Newton method in phases II and III of the workflow.

Before we can begin solving these optimization problems, we require $N_s$ magnetic surfaces in the volume of the device parametrized in Boozer angles.  
The initial surfaces are obtained by starting with the optimized magnetic axis obtained during Phase I.  Then, a first order near-axis expansion is used to find the surface geometry in the neighborhood of the axis with minor radius $r = \SI{0.05}{\meter}$.  That is, the surface degrees of freedom $\mathbf s_1$ are found such that we have
\begin{equation} \label{eq:near-axis-surface}
    \bm \Sigma_{1}(\mathbf s_1, \varphi, \theta) = \bm \Gamma^{\text{axis}}(\varphi) + rX_1(\varphi, \theta)\mathbf{n}(\varphi) + rY_1( \varphi, \theta)\mathbf{b}(\varphi),
\end{equation}
where $\mathbf n, \mathbf b$ are the normal and binormal vectors associated to the Frenet frame of the magnetic axis, $X_1, Y_1$ are defined in \citep{landreman_sengupta_plunk_2019, pyqsc}, and $r$ is a radial distance from the axis.
The remaining $N_s-1$ surfaces with lower aspect ratio are computed using a continuation procedure in the surface label, e.g., volume enclosed by the surface.
During phase III, we already have BoozerLS surfaces available from phase II and those are used as initial guesses for the BoozerExact surfaces.  BoozerLS surfaces are typically very good initial guesses for the BoozerExact solve.

\subsection{Illustration of phase II and III} 
In this section, we give more details on the final two phases of the coil design algorithm to illustrate how it works.
The configurations obtained from the near-axis formulation (phase I) only attempt to find coils that target quasisymmetry on the magnetic axis.  
The formulation does not control anything about the magnetic field away from the axis, and as a result, magnetic surfaces may not exist on a large volume \citep{Lee_2023}.
BoozerLS surfaces (phase II) are robust and capable of healing generalized chaos \citep{surfaceopt2}, though it can be computationally expensive.
Finally, the BoozerExact (phase III) surfaces polish the coil set for precise quasi-axisymmetry and are computationally much cheaper.

The algorithm for phase II is given in Algorithm \ref{alg:alg1}.
The first step of the algorithm is to use the magnetic axis obtained from phase I to compute an approximate magnetic surface from \eqref{eq:near-axis-surface}.
Then, the optimization problem \eqref{eq:optprob2} is solved, but we do not attempt to fully converge the coil sets and only complete a total of 300 iterations of BFGS.
Each time BFGS is restarted, an attempt is made to increase the number of Fourier modes used to represent the surface.
Starting from $m_\text{pol},n_\text{tor}=2$, we try to compute a BoozerLS surface.  If the solve succeeds, we use the surface as an initial guess and increase $m_\text{pol}, n_\text{tor}$ by 1.
 If the surface solve fails after increasing the surface degree, then the degree is reverted to the previous one.  If the solve succeeds, the continuation in degree continues until $m_\text{pol},n_\text{tor}=4$.
Progressively increasing the number of Fourier modes makes the procedure more robust, as we find self-intersecting surfaces are less likely to occur.

\begin{figure}
    \centering
    \includegraphics[width=\textwidth]{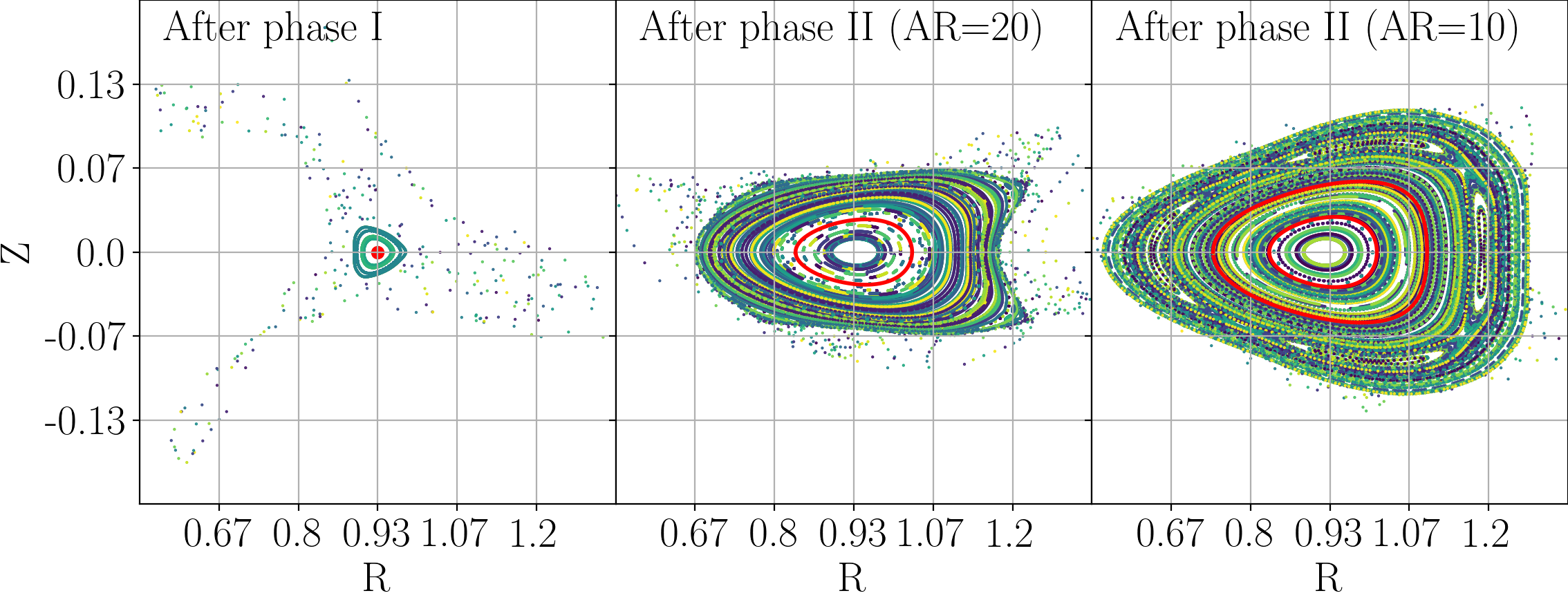}
    \caption{Poincar\'e plots at $\phi=0$ after phase I, and II for aspect ratio $AR=20, 10$.  
    The point where the magnetic axis intersects the $\phi=0$ plane corresponds to the red dot in the first panel.  
    Cross sections of the surfaces on which QA is optimized are red in the second and third panels.
    }
    \label{fig:chaoshealing}
\end{figure}

The weight $w_r$ is chosen such that the Boozer residual term is an order of magnitude larger than the non-quasisymmetry penalty.
This term favors nested flux surfaces and improves the robustness of the optimization algorithm.
However, it also competes with the quasi-axisymmetry error and so the quality of quasisymmetry obtained at this stage can be limited.
Since the stellarators obtained here are not fully converged and strongly depend on the value of $w$, we do not analyze the physics properties of the stellarators here yet.
In Figure \ref{fig:chaoshealing}, we provide Poincar\'e plots of a device just after the near-axis optimization, and then as BoozerLS surfaces are added to optimize for nested flux surfaces.
 The first panel shows that the near-axis optimization algorithm does not guarantee nested surfaces far away from the magnetic axis.  
The final two panels illustrate that as the surfaces are added, the volume with nested flux surfaces increases.

\begin{algorithm}
\caption{Phase II: BoozerLS}\label{alg:alg1}
\begin{algorithmic}
\State compute an initial BoozerLS surface with aspect ratio 20 from near-axis formulas \eqref{eq:near-axis-surface}. 
\For{aspect ratio (AR) in $\{20, 10, 6.66, 5, 4, 3.33, 2.85\}$}
    \For{$i$ in $1,2,3$}
        \State attempt to increase the degree of the BoozerLS surfaces up to $m_{\text{pol}}, n_{\text{tor}} \leq 4$
        \State solve \eqref{eq:optprob2} on surfaces with a fixed budget of iterations. \Comment{Island and chaos healing}
    \EndFor
    \State add another BoozerLS surface with the next smallest AR.
\EndFor
\end{algorithmic}
\end{algorithm}

\begin{algorithm}
\caption{Phase III: BoozerExact}\label{alg:alg2}
\begin{algorithmic}

\For{$i$ in $1,2,3,4$}
    \State keep rational surfaces as BoozerLS, otherwise convert to BoozerExact.
    \State attempt to increase the degree of the BoozerLS surfaces up to $m_{\text{pol}}, n_{\text{tor}} \leq 4$
    \State attempt to increase the degree of the BoozerExact surfaces up to $m_{\text{pol}}, n_{\text{tor}} \leq 10$
    \State solve \eqref{eq:optprob2} on surfaces with a fixed budget of iterations. \Comment{Polish}
\EndFor

\end{algorithmic}
\end{algorithm}

After phase II terminates, we proceed to phase III where we optimize for precise quasisymmetry.
It can be shown that island width scales like $\sqrt{|B_{n,m}|/m\iota'}$ \citep{10.1063/1.1377614}, where $\iota'$ is the magnetic shear.  Thus, for large enough $m$, the island width should be small.  Informed by this estimate, we retain the BoozerLS formulation when the rotational transform on the surface is within 1\% of a rotational transform of the form $\iota = n_{\text{fp}}/m, 2n_{\text{fp}}/m$ for $m=1, \hdots, 15$. 
During this final phase, the weight $w$ is set to zero on BoozerExact surfaces, but it maintained for BoozerLS surfaces in the neighborhood of low-order rationals.  
At the start of every BFGS run, we attempt to increase the degree of the surface representation as we did during Phase II, up to $m_{\text{pol}},n_{\text{tor}}=10$ for BoozerExact surfaces, and 4 for the remaining BoozerLS surfaces.
When all surfaces are of type BoozerExact, we cap the total number of BFGS iterations to 20,000.
When there is a surface of type BoozerLS, we cap the total number of BFGS iterations 4,000.
The algorithm for phase III is given in Algorithm \ref{alg:alg2}.

During these last two phases of coil optimization, we also increase the number of Fourier harmonics in the coils to $N_{f,c}=16$, because using too few Fourier modes in the coils can limit the attainable quality of quasisymmetry.

\subsection{Comparison of devices from TuRBO and naive globalization after phase III}
\begin{figure}
    \centering
    \includegraphics[height=0.2\textheight]{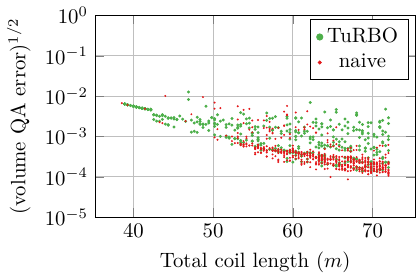}
    \caption[]{Taking the stellarators from Figure \ref{fig:l2g_axis}, and using them as initial guesses for the volume QA optimization phase (Section \ref{sec:opt2}), we obtain the stellarators plotted above.  
    So that the data sets can be more easily distinguished, we use smaller marker sizes for devices corresponding to the naive algorithm.}
    \label{fig:l2g_surface}
\end{figure}

After the full workflow is completed (globalization, then phases I, II, then III), we compare the devices found depending on whether TuRBO or naive globalization approaches were applied just before phase I.  
In Figure \ref{fig:l2g_surface}, we plot volume quasisymmetry error for configurations found for both globalization techniques.  
It appears that the highly favorable devices discovered by TuRBO do not necessarily translate to devices that have comparatively favorable volume quasisymmetry. 
We also find the TuRBO devices resulted in many more configurations for shorter coils than the ones found by the naive approach, though there is a larger spread of device performance.
The Design \ref{fig:l2g_axis}.B does not persist upon optimizing for volume quasisymmetry, which we suspect is because the coils are too close to the magnetic axis ($\SI{16.5}{\centi\meter}$).
As a result, when the surface generated by the near-axis formulas \eqref{eq:near-axis-surface} is used at the start of phase II, the BoozerLS surface solve in \eqref{eq:optprob2:2} does not converge and the optimization cannot proceed.
The devices discovered by TuRBO also do not appear to outperform those found by the naive approach.  Nevertheless, we emphasize that this may only be because the globalization is performed on the near-axis problem just before phase I and \textit{not} directly to the volume QA optimization, just before phase II.
If globalization is applied directly to the BoozerLS phase of the workflow, it is possible that new and interesting devices could still be discovered.
Finally, the magnitude and form of the perturbation $\epsilon$ used in the naive approach was determined after somewhat lengthy tuning (Section \ref{sec:g2l}).
There are much fewer ad-hoc choices in the bounding box approach used by TuRBO.

\subsection{Computational details}
The run time of each phase can vary depending on the design targets, as the difficulty of the optimization problems strongly depends on the total coil length used by the device.
In previous work, we used various levels of parallelism simultaneously such as MPI, OpenMP, and SIMD.
In this work, we take the approach of only allocating one core per stellarator and only use SIMD parallelism locally to a core when possible.
The full workflow for a single device takes on the order of a day or two on a single core.  However, this duration can vary substantially depending on how accurately each individual optimization problem is solved or if more cores are used per problem.

\section{QUASR: the Quasisymmetric Stellarator Repository}\label{sec:volume_database}
The algorithms presented in the previous sections have culminated in the comprehensive database that we detail in this section.  Before discussing the stellarators in the database, we would like to repeat some of the database's limitations:
\begin{itemize}
    \item The database only contains curl-free, stellarator symmetric magnetic fields with optimized quasi-axisymmetry. Note that our algorithms are generic however, and may also be applied to other flavors of quasisymmetry as well.
    \item Our goal was both to produce a large data set of stellarators and to visualize the trade-off of target physics characteristics.  Given that this approach might find stellarators that do not lie on the Pareto front, it may not be the most computationally efficient.  Approaches based on continuation might reduce this computational cost \citep{bindel2023understanding}, at the expense of some parallelism.
    \item The algorithm might have discovered the same device multiple times.
Three possible ways that this can occur are as follows.  Given a local minimum of \eqref{eq:optprob1} or \eqref{eq:optprob2}, a visually distinct local minimum can be found by reflecting the device about the $XY$ plane, i.e., applying the transformation $-Z \to Z$.   Another visually distinct local minimum can also be found by rotating the device by a half-period.
 We have also noticed that local minima can lie in tricky valleys of the objective.  
In particular, initially the gradient-based optimizer makes a lot of progress, and does so quickly.  But after a few thousand iterations, progress slows, especially for devices that use longer coil lengths.
Therefore, it might also be that visually distinct local minima will merge with one another after more iterations, but with only marginal reduction of the objective.  
This is the price paid for an extensive scan.  %
\end{itemize}
Due to the computational expense of generating these devices, we also include in the database the ones discovered after executing previous versions of the workflow described above.
In the following sections, we compare devices in the database to ones discussed previously in the literature, highlight a device with rotational transform profiles that pass through low-order rationals, and perform a couple of trade-off analyses of the devices.  
We look at how accurately quasisymmetry can be attained when the total coil length, number of coils per half period, number of field periods, device aspect ratio, and target mean rotational transform are varied.
In addition, we examine the relationship between elongation and quality of quasisymmetry.
Due to the multiple possible analyses that might be performed, we only scratch the surface here.

\subsection{Comparison to previous devices}
\begin{figure}
    \centering
    \includegraphics[height=0.3\textheight]{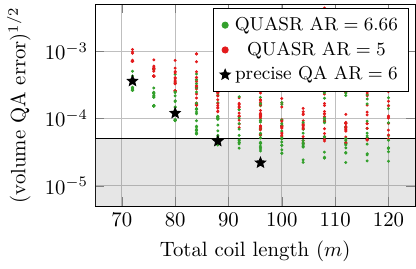}
    \caption{Comparison of devices from QUASR with aspect ratio $6.66$ and $5$, target mean rotational transform $\iota=0.4$, and the precise QA coil sets of \citep{surfaceopt1} with aspect ratio $6$ and mean rotational transform $0.42$.
    All devices have $n_{\text{fp}}=2$ and $n_{\text{coils per hp}}=4$ here.
    The grey region corresponds to the Earth's background magnetic field of $\SI{50}{\micro \tesla}$, and volume QA error refers to the mean non-QA ratio in \eqref{eq:objf4}.
    }
    \label{fig:comparison}
\end{figure}

As a first examination, we select the devices in our database that have the closest design targets to previous configurations computed in \citep{surfaceopt1}, where devices with aspect ratio 6, $\iota=0.42$, $n_{\text{coils per hp}}=4$, total coil length $\SI{72}{\meter}, \SI{80}{\meter}, \SI{88}{\meter}, \SI{96}{\meter}$, and $n_{\text{fp}}=2$.
The devices in QUASR that are closest to these devices have the same number of coils, aspect ratio $6.66$ or $5$, and target mean rotational transform $0.4$.
In Figure \ref{fig:comparison}, these devices are compared, where we observe that the QUASR devices perform comparably, though notably the precise QA device with length $\SI{96}{\meter}$ is better.
This might be because we only allow a total of 20,000 iterations of BFGS during phase III, while in generating the precise QA coil sets, more than double the number of iterations were allowed.

\subsection{A stellarator with a rotational transform profile passing through low-order rational}
\begin{figure}
    \centering
    \begin{tikzpicture}
        \node (A) at (0,0)    {  \includegraphics[width=\textwidth]{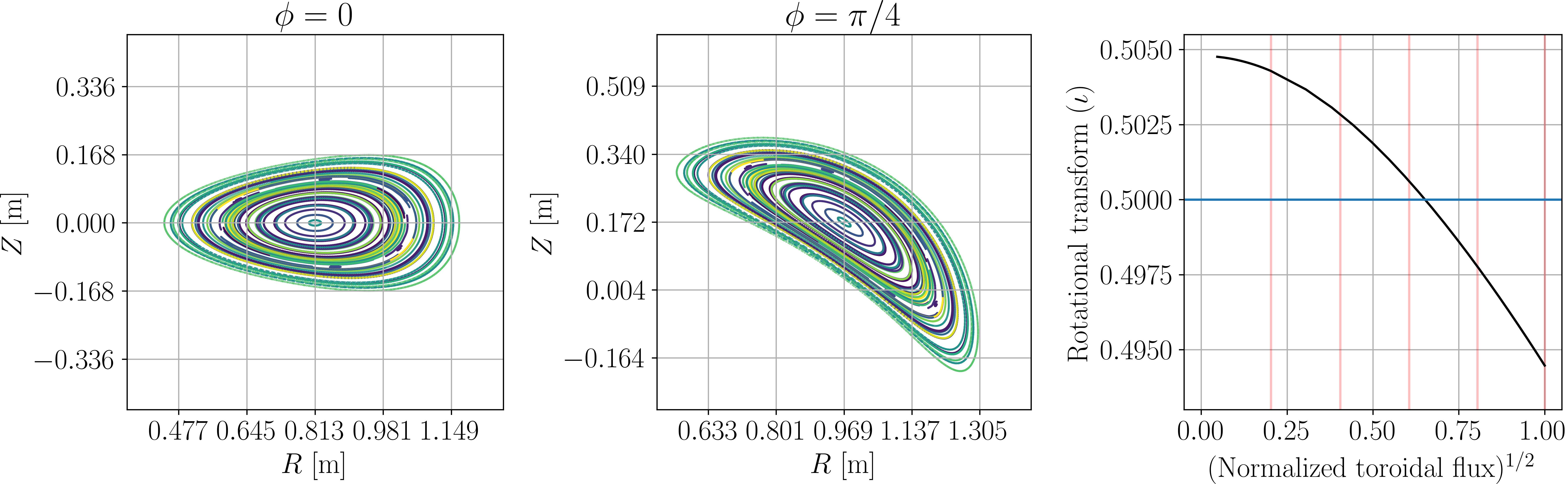}};
    \end{tikzpicture}
    \caption{Poincar\'e plots and rotational transform profiles of a device with aspect ratio 4, mean rotational transform 0.5, and $n_{\text{fp}}=2$.
    We visually could not identify trajectories associated to islands.
    The horizontal blue lines on the rotational transform profiles indicate the low order rational $\iota=1/2$.
    The vertical red lines on the rotational transform profiles indicate the normalized toroidal flux label on which quasisymmetry was optimized.}
    \label{fig:interesting_configs}
\end{figure}

Consider a devices with two field periods, $n_{\text{coils per hp}}=3, \iota_{\text{target}}=0.5$ and aspect ratio 4.  Since the target mean rotational transform is a low-order rational, we would expect there to be a strong likelihood of an island chain somewhere in the toroidal volume.  However, the island chains are small, as shown in Figure \ref{fig:interesting_configs}.

\subsection{Impact of design targets on attainable quasisymmetry}
\begin{figure}
    \centering
    \includegraphics[width=\textwidth]{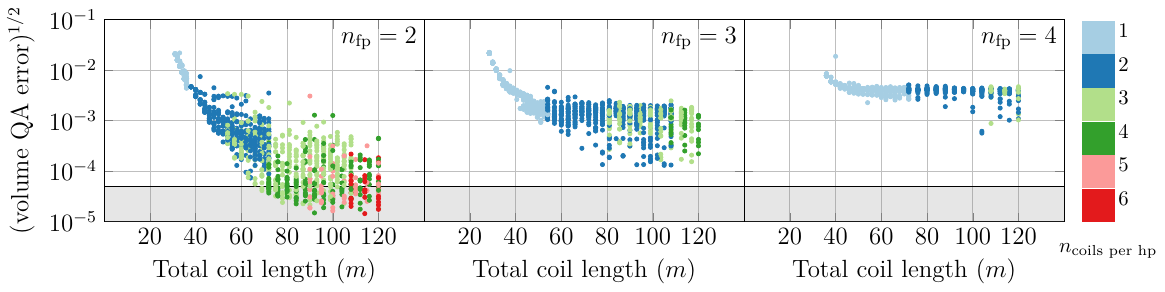}
    \caption{Devices on which quasisymmetry is optimized on a surface of aspect ratio 20, with a target rotational transform $\iota=0.5$.
    The grey region corresponds to the Earth's background magnetic field of $\SI{50}{\micro \tesla}$, and volume QA error refers to the mean non-QA ratio in \eqref{eq:objf4}.
    }
    \label{fig:full_surface_database}
\end{figure}

In Figure \ref{fig:full_surface_database}, we plot the volume QA error with respect to coil length for various values of $n_{\text{fp}}=2, 3, 4$, when optimizing for QA on a single surface of aspect ratio 20.  Since this is such a high aspect ratio surface, this set up is a close approximation to the near-axis problem in Section \ref{sec:opt1}, though the objective is asking for more out of the magnetic field. 
A notable difference with the near-axis results is that increasing the number of field periods does not appear to improve the attainable on-axis quasisymmetry, as is observed in Figure \ref{fig:full_axis_database}.
One possible reason for this is that the near-axis design problem \eqref{eq:optprob1} only optimizes for quasisymmetry to first-order and introducing second-order near-axis penalties might explain this discrepancy.
There do appear to be some highly performant devices that are outliers, so it might also be that our algorithm is just not discovering those solution branches. 
There is also a clear preference for $n_{\text{fp}}=2$ stellarators as the algorithm has difficulty finding devices with precise quasisymmetry for $n_{\text{fp}}=3,4$.  
Devices with $n_{\text{fp}}=1,5$ are not shown in the figure, but in both cases, the algorithm had varying difficulty finding devices with precise quasi-axisymmetry.

\begin{figure}
    \centering
    \includegraphics[width=\textwidth]{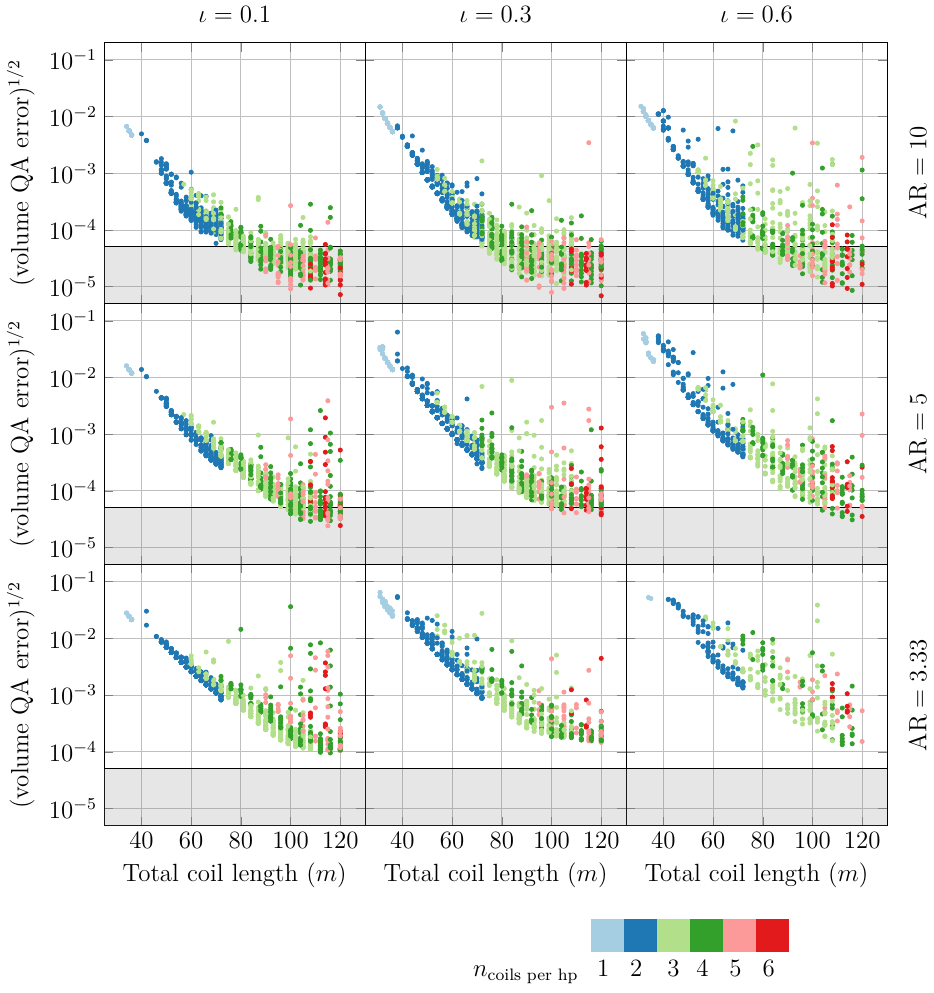}
    \caption{Trade-off between quality of quasisymmetry and total coil length for various device aspect ratios, and target mean rotational transform when $n_{\text{fp}}=2$.
     The grey region corresponds to the Earth's background magnetic field of $\SI{50}{\micro \tesla}$, and volume QA error refers to the mean non-QA ratio in \eqref{eq:objf4}.
    }
    \label{fig:tradeoff3D}
\end{figure}

We also examine the trade-off between the quality of quasisymmetry, total coil length, and device aspect ratio in Figure \ref{fig:tradeoff3D}.  Increasing the target mean rotational transform does not appear to greatly affect the quality of attainable quasisymmetry, while as expected decreasing the device aspect ratio appears to have a much larger negative impact.

\subsection{Maximum elongation}
Finally, we study the values of elongation appearing in the database when $\iota_{\text{target}}=0.6$ and aspect ratio 10.
High elongation increases the ratio of surface area to volume of the magnetic surfaces, moreover, it can increase bunching of the flux surfaces and this can have negative implications on stability \citep{goodman2022constructing}.
For each stellarator in our database, we compute the maximum elongation in the device by computing a surface with aspect ratio 80 in the neighborhood of the magnetic axis.  Then, we compute $N=10$ cross sections of that surface at cylindrical angles $\phi=\pi(i+1/2)/n_{\text{fp}}/N \text{ for } i= 0, \hdots, N-1$.  For each cross section, we fit an ellipse in a least squares sense \citep{halir1998numerically} and compute the ratio of the ellipse's major to minor axes. The value reported here is the maximum ratio observed at these cross sections.

\begin{figure}
    \centering
    \includegraphics[width=\textwidth]{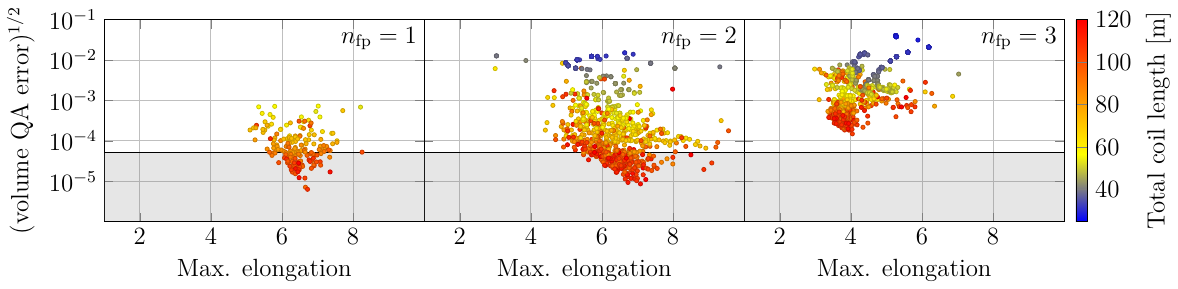}
    \caption{The volume quasisymmetry error with respect to the maximum elongation on a very high aspect ratio surface in the neighborhood of the magnetic axis, for various values of $n_{\text{fp}}$.  For all configurations plotted, $\iota_{\text{target}}=0.6$, where the aspect ratio of the outermost surface is 10. The grey region corresponds to the Earth's background magnetic field of $\SI{50}{\micro \tesla}$, and volume QA error refers to the mean non-QA ratio in \eqref{eq:objf4}.
    }
    \label{fig:max_elongation}
\end{figure}

We did not target any particular value of elongation but nevertheless there do appear to be favored values (Figure \ref{fig:max_elongation}).
For $n_{\text{fp}}=1$ and $2$, the lowest values of quasisymmetry error occur for an elongation around 7, while for $n_{\text{fp}}=3$ an elongation of 4 is favored.
It is unclear whether this picture would change if a stage one optimization for QA were done instead, without coils.
In \citep{goodman2022constructing}, various QI stellarator designs were proposed, where a maximum elongation below approximately 6 was targeted.
The stellarators in QUASR here illustrate that a good approximation of QA can also be found when requiring a maximum elongation below 6 too.
We also observe that as expected, the lower quasisymmetry errors occur for longer coil lengths.

\section{Conclusions}
We have proposed a direct stellarator coil design algorithm that is globalized using the TuRBO optimization algorithm, where box constraints on anchor points of the coils are applied.
The algorithm combines three direct coil optimization algorithms, and has allowed the construction of a large database of approximately 200,000 vacuum field stellarators for various design targets, e.g., aspect ratio, $n_{\text{fp}}$, $n_{\text{coils per hp}}$, rotational transform, and total coil length.
Using the database, we have examined the trade-off between accuracy of quasi-axisymmetry, total coil length, rotational transform, and aspect ratio of the device.

Since the techniques in this work are quite general, there are many other directions that we would like to explore.
Applying all of these approaches to other flavors of quasisymmetry, such as quasi-helical symmetry is the next logical step.
Adding windowpane coils and other coil geometries, e.g., helical coils, to the database would enrich the database.
There is still room to improve the globalization algorithms and one possibility is to directly globalize the BoozerLS phase of the optimization.

Finally, we have only scratched the surface of possible physics analyses of the data set.  Since it is publicly available, we hope that the stellarator community might explore it further.

\section{Data availability statement}
\begin{enumerate}
    \item Scripts that wrap Phase I with the TuRBO globalization are available at \url{https://github.com/andrewgiuliani/Global-Direct-Coil-Optimization-I}.
    \item Scripts for Phase II and III are available in SIMSOPT \citep{Landreman2021} and \url{https://github.com/andrewgiuliani/Global-Direct-Coil-Optimization-II}.
    \item The entire set of stellarators in VMEC and SIMSOPT formats is archived on Zenodo at \url{https://doi.org/10.5281/zenodo.10050656}.
    \item Jeff Soules has written a web application for navigating the data set, hosted at \url{https://quasr.flatironinstitute.org/}. 
\end{enumerate} 

\section{Acknowledgements}
The author would like to thank the Flatiron Institute's Scientific Computing Core for their support, Georg Stadler, Gabriel Provencher Langlois, and Rogerio Jorge for the helpful discussions, and Misha Padidar for the tips to get set up with TuRBO.
We also would like to thank Rogerio Jorge for linking PyQSC to SIMSOPT for obtaining initial surface guesses at the start of Phase II in the workflow.
Finally, the author would like to thank Jeff Soules for writing the web application to navigate the database.

\section{Declaration of interests}
The authors report no conflict of interest.

\section{Author ORCID}
Andrew Giuliani: 0000-0002-4388-2782

\bibliographystyle{jpp}
\bibliography{main_revision_2}

\end{document}